\documentclass[twocolumn,showpacs,preprintnumbers,amsmath,amssymb]{revtex4}
\usepackage{graphicx}% Include figure files
\usepackage{dcolumn}% Align table columns on decimal point
\usepackage{bm}% bold math
\usepackage{amsmath}
\usepackage{amsfonts}% mathbb fonts
\usepackage{mathrsfs}
\usepackage{hyperref}
\usepackage{amssymb}

\begin{document}

\title{Shannon dimensionality of quantum channels and its application to photon entanglement}
\author{J.B. Pors}
\email{pors@molphys.leidenuniv.nl} \affiliation{Huygens
Laboratory, Leiden University, P.O.\ Box 9504, 2300 RA Leiden, The
Netherlands}
\author{S.S.R. Oemrawsingh}
\altaffiliation{present address: Department of Physics, University
of California, Santa Barbara, California 93106, USA}
\author{A. Aiello}
\affiliation{Huygens Laboratory, Leiden University, P.O.\ Box
9504, 2300 RA Leiden, The Netherlands}
\author{M.P. van Exter}
\affiliation{Huygens Laboratory, Leiden University, P.O.\ Box
9504, 2300 RA Leiden, The Netherlands}
\author{E.R. Eliel}
\affiliation{Huygens Laboratory, Leiden University, P.O.\ Box
9504, 2300 RA Leiden, The Netherlands}
\author{G. W. 't Hooft}
\affiliation{Huygens Laboratory, Leiden University, P.O.\ Box
9504, 2300 RA Leiden, The Netherlands}
\author{J.P. Woerdman}
\affiliation{Huygens Laboratory, Leiden University, P.O.\ Box
9504, 2300 RA Leiden, The Netherlands}

\begin{abstract}

We introduce the concept of Shannon dimensionality $D$ as a new
way to quantify bipartite entanglement as measured in an
experiment. This is applied to orbital-angular-momentum
entanglement of two photons, using two state analyzers composed of
a rotatable angular-sector phase plate that is lens-coupled to a
single-mode fiber. We can deduce the value of $D$ directly from
the observed two-photon coincidence fringe. In our experiment, $D$
varies between 2 and 6, depending on the experimental conditions.
We predict how the Shannon dimensionality evolves when the number
of angular sectors imprinted in the phase plate is increased and
anticipate that $D\simeq 50$ is experimentally within reach.

\end{abstract}

\maketitle

Photons can be entangled in various degrees of freedom. The most
extensively studied variety involves the polarization degrees of
freedom, of which there are inherently two per photon. In a
typical EPR-Bell type experiment, the state analyzers are
polarizers, and when their relative orientation is scanned, this
gives rise to a sinusoidal coincidence fringe \cite{Aspect}. This
particular shape is characteristic of the two-dimensional nature
of polarization entanglement.

Recently, much attention has been drawn to bipartite entanglement
involving more than two degrees of freedom. With increasing
dimensionality, quantum entanglement becomes correspondingly
richer. High-dimensional entanglement is predicted to violate
locality more strongly and to show more resilience to noise
\cite{Collins,Kaszlikowski}. From an applications perspective, it
holds promise for implementing larger alphabets in quantum
information, e.g. quantum cryptography
\cite{Bechmann-Pasquinucci}, and for an increased security against
eavesdropping \cite{Zhang}. High-dimensional entanglement can
be studied employing the frequency-time \cite{Riedmatten} or
position-momentum degrees of freedom, the latter having been
demonstrated for both the transverse linear \cite{Neves,
O'Sullivan-Hale} and orbital-angular-momentum degrees of freedom
\cite{Mair, Oemrawsingh2005}.

It is crucial to have a quantifier of the dimensionality of
entanglement as measured in an experiment \cite{Brunner}. In this
Letter we introduce such a quantifier, using concepts from
classical information theory in the spirit of Shannon
\cite{ShannonBook}. We apply these ideas to
orbital-angular-momentum entanglement, inserting appropriate
angular state analyzers in the beamlines of a parametric
down-conversion setup. We have realized a Shannon dimensionality
$2\leq D \leq 6$ and we argue that $D\simeq50$ is within reach.

In classical information theory \cite{ShannonBook}, the number of
independent communication channels of a signal is known as the
Shannon number. The signal being the state of a physical system,
the Shannon number is also referred to as the number of degrees of
freedom, or the number of modes, of that system \cite{Toraldo69,
Gori}. For example, a signal encoded in the polarization degrees
of freedom of a light beam has a Shannon number equal to 2.

When dealing with a bipartite quantum system in an entangled pure
state $|\psi \rangle \in \mathscr{K} = \mathscr{K}_A \otimes
\mathscr{K}_B$, the usual measure of the effective dimensionality
of the Hilbert space in which the state lives, is given by the
Schmidt number $K$ \cite{Karelin}
\begin{equation}\label{Eq.1}
    K = \frac{1}{\mathrm{Tr}_A(\rho_A^2)} =
    \frac{1}{\mathrm{Tr}_B(\rho_B^2)}.
\end{equation}
Here, $\rho_A = \mathrm{Tr}_B(|\psi \rangle \langle \psi |)$ and
$\rho_B= \mathrm{Tr}_A(|\psi \rangle \langle \psi |)$, are the
reduced density matrices representing the states of the two
sub-systems $A \in \mathscr{K}_A$ and $B \in \mathscr{K}_B$,
respectively. Although a system may have infinitely many degrees
of freedom, any actual measurement apparatus has effective access
only to a finite number of them, say $D$. Such a dimensionality
$D$ is referred to as the Shannon number of the measurement
apparatus.

Consider an experiment measuring correlations between the two
subsystems $A$ and $B$. There are two  measuring apparatuses, say
$\mathscr{P}_A(\alpha)$ and $\mathscr{P}_B(\beta)$, interacting
with subsystems $A$ and $B$, respectively, where $\alpha$ and
$\beta$ label possible settings of the two apparatuses. For a
given setting $\xi \in \{ \alpha, \beta\}$, detector
$\mathscr{P}_X(\xi)$ is represented by the projection operator
$\hat{\Gamma}(\xi) = |X(\xi)\rangle \langle X(\xi)|$, where $X \in
\{A,B\}$, and $|X(\xi)\rangle$ is the state in which the system
$X$ is left after measurement.

If a Von Neumann-type projective measurement is performed, the set
of states $\{ |X(\xi) \rangle \}_\xi$ obtained by varying $\xi$ is
complete and orthonormal, namely
\begin{equation}\label{Eq.2}
 \langle X(\xi) |  X(\xi') \rangle  = \delta_{\xi \xi'}, \qquad \sum_\xi \hat{\Gamma}(\xi) = \hat{1},
\end{equation}
where the measurement operators $\hat{\Gamma}(\xi)$ are Hermitean
and idempotent. The number of these operators is equal to the
dimension of the Hilbert space of the measured quantum system
\cite{Brandt}. However, in many situations non-orthogonal
measurements are made and Eqs. (\ref{Eq.2}) do not hold
\cite{Thew}. In this case, the number of projection operators
$\hat{\Gamma}(\xi)$ does not give the dimension of the Hilbert
space of the measured system, and a new criterion must be
introduced.

Let us therefore consider finite-dimensional systems, say $\dim
(\mathscr{K}_X) = L$, and rewrite Eq. (\ref{Eq.2}) for the case of
non-orthogonal measurements as
\begin{equation}\label{Eq.3}
 \langle X(\xi) |  X(\xi') \rangle  = g_{\xi \xi'}, \qquad  \sum_\xi \hat{\Gamma}(\xi) = \hat{\gamma},
\end{equation}
where $G=[g_{\xi \xi'}]$ is a matrix of size $L \times L$, and
$\hat{\gamma}$ is an Hermitean operator. The eigenvalues
$\gamma_l$ of $\hat{\gamma}$ give the detector's `sensitivity' to
the corresponding eigenmodes. In general, a detector will not be
equally sensitive to all eigenmodes and some $\gamma_l$ are
substantially larger than others. The \textit{effective}
dimensionality $D \leq L$ of the Hilbert space $\mathscr{D}$ where
the \textit{measured} system lives can be quantified as the
Hilbert-Schmidt norm of the eigenvalue distribution \cite{Pors}
\begin{equation}\label{Eq.4}
    D \equiv \frac{1}{\text{Tr} (\hat{\gamma}^2)} = \frac{1}{\sum_l \gamma^2_l}.
\end{equation}
This dimensionality should be interpreted as the effective Shannon
number of information channels \cite{ShannonBook, Toraldo69}.

The isomorphism of Eq. (\ref{Eq.1}) and Eq. (\ref{Eq.4}) suggests
a relation between the Schmidt number $K$ and the Shannon
dimensionality $D$. The nature of such relation becomes clear if
one notes that since the operators $\hat{\Gamma}(\xi)$ are
Hermitian and positive semidefinite, the operator $\hat{\gamma}$
may be interpreted as a density matrix acting in $\mathscr{K}_X$
\cite{BengtssonBook}. Thus, if we think of $\hat{\gamma}$ as a
reduced density matrix of a bipartite system, then $K$ and $D$ are
formally the same. However, it is important to note that while $K$
furnishes the dimensionality of the \textit{generated}
entanglement, $D$ gives the effective dimensionality of the space
$\mathscr{D}$ that can \textit{potentially} be \textit{probed} and
it is a property of the projection apparatus only. The
dimensionality of the \textit{measured} entanglement is a joint
property of the generated system and analyzers, but simply amounts
to $D$ as long as $\mathscr{K}\supset\mathscr{D}$.

%%%%%%%%%%%%%%%%%%%%%%%%%%%%%%%%%%%%%%%%%%%%%%%%%%%%%%%%%%%%%%%%%%%%%%%%%%%%%%%%%%%%%%%%%

\begin{figure}[!hbr]
\includegraphics[angle=0,width=8.3truecm]{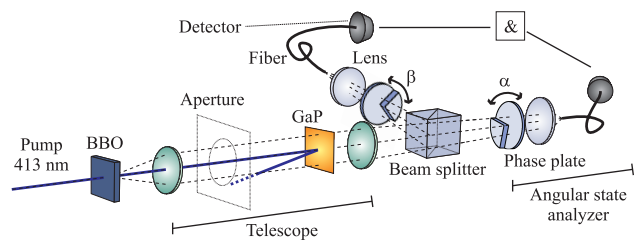}
\caption{\label{fig:1} (color online). Experimental setup.
Orbital-angular-momentum entangled photons are emitted at 826 nm
by a BBO crystal, cut for Type-I collinear phase matching. A thin
GaP wafer serves to eliminate the pump beam. The two-photon field
can be clipped with an aperture. The twin-photons are spatially
separated by a beam splitter and imaged on the angular phase
plates ($f_2=4f_1=40~$cm). Just behind the phase plates, the
frequency-degenerate photons are selected by interference filters
(not shown), centered around 826 nm with a 10 nm width. The phase
plates (shown are quarter-sector plates) are oriented at angles
$\alpha$ and $\beta$, and photon counts are rendered by a
coincidence circuit. }
\end{figure}

Next, we apply our formal theory to an experiment on
orbital-angular-momentum entanglement of two photons, in order to
illustrate how detector characteristics bound the measured
entanglement to an effective Shannon dimensionality $D$, while
probing a generated state with Schmidt number $K \gg D$ (and
$\mathscr{K}\supset\mathscr{D}$). Our experimental setup is
depicted in Figure \ref{fig:1}. Pumping a BBO non-linear crystal
with a 150 mW Kr$^+$ laser beam at $\lambda=413$ nm, we produce
spatially entangled photons by means of spontaneous parametric
down conversion. The state we generate is of the form
$|\Psi\rangle = \sum_l \sqrt{\lambda_l} |l\rangle |-l\rangle$,
where $|l\rangle$ denotes the orbital angular momentum eigenmode
of order $l$: $\langle \phi | l \rangle = \exp(i l
\phi)/\sqrt{2\pi}$, with $\phi$ the azimuthal angle
\cite{Walborn}. Employing Type-I collinear phase matching, we
collect the full emission cone and with the experimental
parameters of our setup (beam half-waist at the position of the
crystal $w_\text{0} = 250~ \mu$m and crystal length 1 mm) we
obtain an azimuthal Schmidt number $K\simeq31$ \cite{Law,
Exter_SpaFilt, Exter_ModCount}. The twin photons are spatially
separated by means of a non-polarizing beam splitter.

Each arm of the setup contains an angular state analyzer, composed
of an angular phase plate that is lens-coupled to a single-mode
fiber (see Fig. \ref{fig:1}) \cite{Oemrawsingh2006}. The angular
phase plates carry a purely azimuthal variation of the optical
thickness. As in polarization entanglement \cite{Aspect}, the
phase plates are rotated around their normals and the photon
coincidence rate is recorded as a function of their independent
orientations \cite{Oemrawsingh2005}.

The combined detection state of the two angular-phase-plate
analyzers, each acting locally, can be expressed as
\begin{equation}
    \label{Eq.5}
    |A(\alpha)\rangle \otimes |B(\beta)\rangle = \left(\sum_l \sqrt{\gamma_l} |l\rangle e^{i l \alpha}\right)_A \otimes \left(\sum_{l} \sqrt{\gamma_{l}} |l\rangle e^{i
    l \beta}\right)_B,
\end{equation}
where $\alpha$ and $\beta$ denote the orientations of the two
phase plates, respectively \cite{Footnote1}. The complex expansion
coefficients $\sqrt{\gamma_l}$ are fixed by the physical profile
of the angular phase plate and obey the normalization condition
$\sum_l |\gamma_l|=1$. In general, the detection state constitutes
a non-uniform superposition of orbital angular momentum modes.
When the angular phase plates are rotated over $\alpha$ or
$\beta$, all modes in the superposition rephase with respect to
each other, yielding a set of detection states of the type Eq.
(\ref{Eq.3}). The effective Shannon dimensionality that is so
probed is given by Eq. (\ref{Eq.4}). It is the average number of
modes captured by an analyzer when its phase plate is rotated over
$360^\circ$.

As we have recently shown in Ref. \cite{Pors}, the Shannon
dimensionality is straightforwardly deduced from the shape of the
experimental coincidence curve; it is the inverse of the area
underneath the peak-normalized coincidence fringe, obtained when
rotating one of the phase plates.

\begin{figure}[!hbr]
\includegraphics[angle=0,width=7.8truecm]{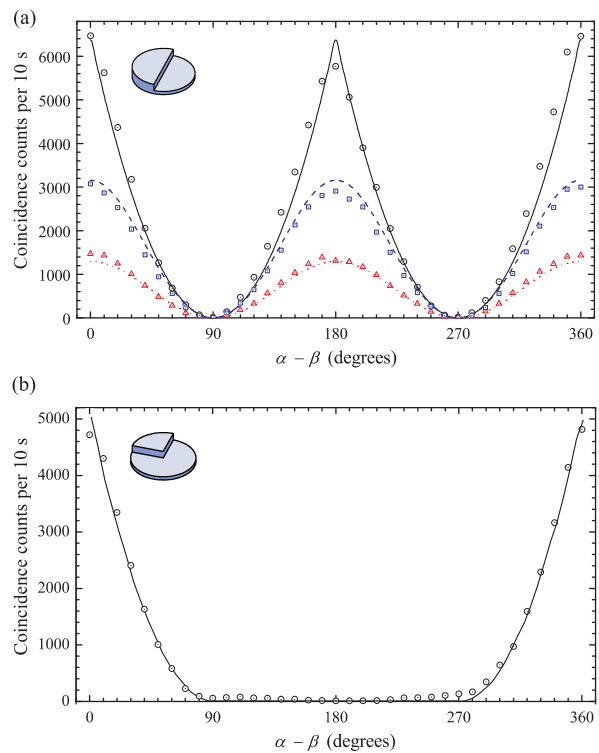}
\caption{\label{fig:2} (color online). Coincidence count rate vs.
the relative orientation of the two state analyzers. Points denote
experimental data, the curves are theoretical predictions. (a)
Half-sector plate. The parabolic fringe (circles) is a signature
of a dimensionality larger than two: we find $D=3.0$. Truncating
the number of modes, by closing the aperture, gradually reduces
the parabola into a sine of dimensionality 2.0 (triangles). (b)
Quarter-sector plate. The piece-wise parabolic fringe yields an
experimental dimensionality of 5.8 (circles), where theory
predicts $D=6$.}
\end{figure}

In our experiment, we have used angular-sector phase plates; these
have a single arc sector, characterized by the angle $\delta$,
whose optical thickness is $\lambda/2$ greater than that of the
remainder of the plate \cite{Oemrawsingh2006}. The part of the
field that crosses this sector thus flips sign. The phase plates
are manufactured from fused-quartz plane-parallel plates, having a
wedge angle of $0.25"$. They are processed by a combination of
photolithography, wet etching, deposition and lift-off, resulting
in a well-defined mesa structure, with a transition region that is
typically $20~\mu$m wide. The insets of Figure \ref{fig:2} show
two such plates; a half-sector plate ($\delta=\pi$) consisting of
two equal halves that are phase shifted by $\pi$; and a
quarter-sector plate ($\delta=\pi/2$) having one quadrant
$\pi$-phase shifted with respect to the remainder of the plate.

For state analyzers that are equipped with such sector phase
plates, the Shannon dimensionality is given by \cite{Pors}
\begin{equation}
\label{Eq.6}
D(\delta) = \left\{%
\begin{array}{ll}
   \left[1 - 4\frac{\delta}{\pi} + 6\left(\frac{\delta}{\pi}\right)^2 - \frac{8}{3}\left(\frac{\delta}{\pi}\right)^3 \right]^{-1}, & \delta \in [0 ,\pi], \\
   D(2 \pi -\delta), & \delta \in [\pi,
2\pi]. \\
\end{array}
\right.
\end{equation}
For $\delta=0$ we find the trivial result $D=1$; a planar plate
does nothing. For $\delta=\pi$, i.e. a state analyzer equipped
with a half-sector plate, we arrive at $D=3$. For an analyzer
equipped with a quarter-sector plate we find $D=6$. This is the
maximum value for a single angular-sector phase plate. We note
that for our setup indeed $K \gg D$.

%%%%%%%%%%%%%%%%%%%%%%%%%%%%%%%%%%%%%%%%%%%%%%%%%%%%%%%%%%%%%%%%%%%%%%%%%%%%%%%%%%%%%%%

In the experiment, we scan one angular-sector phase plate over a
$360^{\circ}$ rotation, the other remaining fixed, and measure the
coincidence rate. In terms of Klyshko's picture of advanced waves
\cite{Klyshko}, valid when $K \gg D$, the resulting shape of the
coincidence curves can be explained in terms of the mode overlap
of the two state analyzers. Figure \ref{fig:2}(a) shows
experimental results obtained with two half-sector plates
$(\delta=\pi)$, having step height of 0.48 $\lambda$. The data
points form a double \textit{parabolic} fringe, consistent with
theory (solid curve). The maxima at $0^\circ$ and $180^\circ$ are
sharply peaked. The zeros of the fringe are very deep; less than
10 counts per 10 seconds. The maximum coincidence rate is of the
order of $6.5\times 10^3$ per 10 seconds, compared to $10^5$
single counts. We verified that the coincidence rate depends on
the \textit{relative} orientation between the two phase plates
only, the fringe visibility being $>99\%$ for all cases studied.
This basis independence is the key aspect of quantum entanglement.
From the area underneath the data we deduce the experimental value
$D=3.0$. Note that a parabolic fringe was also reported in Ref.
\cite{Oemrawsingh2005}, obtained with non-integer spiral phase
plates. We conclude that also in that case $D=3$.

An aperture, positioned inside the telescope, allows us to control
the number of detected modes (see Figure \ref{fig:1}). Because of
the anti-symmetric profile of the half-sector plate, the detection
state contains only odd expansion terms (see Eq. (\ref{Eq.5})) in
a fashion $\gamma_l = \gamma_{-l}$. When the aperture size is
reduced, higher-order orbital-angular-momentum modes are cut off
so that, eventually, only the modes $l=1$ and $l=-1$ survive. We
then expect a sinusoidal fringe, analogous to two-dimensional
polarization entanglement \cite{Aspect}. In the experiment, we
observe that the coincidence curve is gradually transformed from
parabolic to sinusoidal when the aperture gets smaller. Using an
aperture of 600 $\mu$m diameter, we are in an intermediate regime
(squares, $D=2.1$), while using a 400 $\mu$m diaphragm yields a
curve that resembles a sine very well (triangles, $D=2.0$). The
dashed and dotted curve are theoretical predictions.

To achieve $D=6$, we use two quarter-sector plates
$(\delta=\pi/2)$, carrying an edge discontinuity deviating less
than 3\% from $\lambda/2$. The circles in Figure \ref{fig:2}(b)
show our experimental results, revealing a coincidence curve which
is parabolic for $|\alpha-\beta| \leq 90^\circ$ and equal to zero
otherwise, in agreement with theory (solid curve). We find
$D=5.8$, in very good agreement with the expected value of 6
mentioned above.

%%%%%%%%%%%%%%%%%%%%%%%%%%%%%%%%%%%%%%%%%%%%%%%%%%%%%%%%%%%%%%%%%%%%%%%%%%%%%%%%%%%%%%

\begin{figure}[!hbr]
\includegraphics[angle=0,width=7.8truecm]{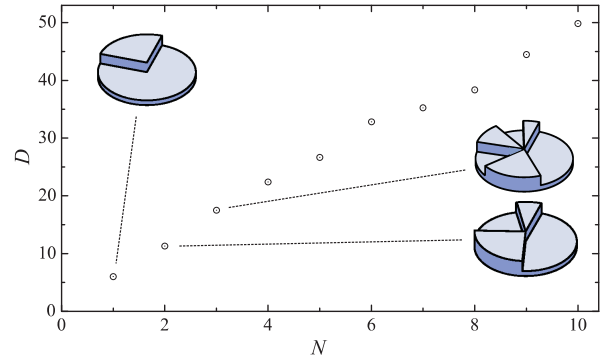}
\caption{\label{fig:3} (color online). Maximum dimensionality that
can be accessed with sector phase plates having $2N$ angular
sectors alternatingly phase shifted by $\pi$. The insets show the
optimized plates for $N=1, N=2$ and $N=3$.}
\end{figure}
The maximum value of the Shannon dimensionality that can be
achieved with a phase plate having but a single sector is $D=6$.
Can one reach higher values of $D$ by using plates with more
sectors? To answer this question, we consider plates with $N$
sectors that are phase shifted by $\pi$ with respect to
interjacent regions. For each choice of sector angles, we
calculate the expansion coefficients $\{\sqrt{\gamma_l}\}$ and,
subsequently, $D$ (see Eq. (\ref{Eq.4}) and (\ref{Eq.5})). Next,
we maximize $D$ by adjusting the sector angles using a Monte-Carlo
random-search algorithm. The result is plotted in Figure
\ref{fig:3}, showing a graph of the maximum value of $D$ versus
the number of mesas $N$. For 10 such sectors, we find $D=49.9$.
The insets show the optimal phase plates for $N=1$ (quarter-sector
plate), $N=2$, and $N=3$.

In conclusion, we have introduced the effective Shannon
dimensionality as a novel quantifier of entanglement as measured
in an actual experiment. We have demonstrated its significance to
the case of two-photon orbital-angular-momentum entanglement.
Using angular-sector phase plates, we have achieved Shannon
dimensionalities up to $D=6$. We anticipate that it is feasible to
probe dimensionalities as high as 50, using multi-sector phase
plates. These can be manufactured by means of photo- or e-beam
lithography as in diffractive-optics technology. Alternatively,
the use of adaptive optical devices, such as spatial light
modulators or micro-mirror arrays, seems promising, particularly
because of their versatility with regard to plate patterns.
However, the ultimate limit to the Shannon dimensionality is
constrained by the angular Schmidt number of the source; using
periodically poled crystals, such as PPKTP, $K\sim100$ is viable
for realistic values of pump-beam waist and crystal length,
without loss of count rates \cite{Fiorentino}.

This work was supported by the Stichting voor Fundamenteel
Onderzoek der Materie.

%\bibliography{bibliography}

\begin{thebibliography}{27}
\bibitem[{Foo()}]{Footnote1}
\bibinfo{note}{The set of OAM eigenmodes constitutes the appropriate basis to
  expand the detection state, for these modes are eigenfunctions of rotation
  over $\alpha$, and the set contains the fiber mode $l=0$.}
\end{thebibliography}

\end{document}